\documentclass[aps,prl,twocolumn]{revtex4}
\tighten

 \usepackage{times}

\begin{document}

\title{Infinitesimal change is a Non-local Operation: Hidden power of Quantum
Entanglement }

\author{Arun K. Pati 
}

\affiliation{Institute of Physics, Sainik School Post, 
Bhubaneswar-751005, Orissa, India}

\date{\today}

\def\ra{\rangle}
\def\la{\langle}
\def\ver{\arrowvert}

\begin{abstract}
We show that the global infinitesimal change in the multi-particle 
pure product 
state gives rise to an entangled state. This suggests that even if there 
is no interaction present between the subsystems, i.e., at each time 
instant the 
state is non-entangled, the tangent vector is typically entangled. 
Since the tangent space vectors tell the state-space vectors how 
to change this implies that quantum entanglement is necessary for 
motion or change in general. This is truly a  `hidden power' of quantum 
entanglement. During quantum computation even though at each time instant 
the state is not entangled, quantum entanglement guides the process of 
computation. This observation applies to multi-particle pure, pseudo-pure 
and  mixed states as well.
\end{abstract}

\maketitle


\vskip 1cm


In the early days, the notion 
of quantum entanglement was much debated concept \cite{epr}. 
Now, quantum entanglement is one of the much studied subject due to
its potential application in information processing \cite{nc}. 
Supplemented by classical 
communication, quantum entanglement can become a resource for very useful and 
exotic information processing tasks.  It is also argued that quantum 
entanglement may play an important role in quantum algorithms \cite{jozsa} 
and in giving extra power to quantum computers \cite{fitz}.

In this paper I make a simple yet an important observation that could 
throw some light on the role of entanglement in quantum evolution and
this in turn may answer the question 
where from the extra power comes for quantum computation \cite{fitz}. 
We show that for 
any multi-particle state if two or more subsystems undergo generic change then 
the infinitesimal change in the pure product state gives 
rise to an entangled state. In the language of differential geometry given any 
manifold of multi-particle quantum states (product or entangled), 
the tangent space to each state is an entangled manifold. 
In other words the tangent vectors are 
typically `entangled' even when the states themselves are not. Since the 
tangent vectors describe the action of the Schroedinger equation, there 
would seem to be some intimate relation between entanglement and the 
physical nature of change itself.
We can say that {\em entanglement is necessary for 
any multi-particle continuous quantum evolution.} 
This I call the  `hidden power' of quantum entanglement. Applied to quantum 
computation, this implies that during computation even though at 
each time instant the state is a product state, 
entanglement is necessary for quantum computation. 
Without entanglement there is no generic change. In other words, 
{\em even non-entangling evolution needs quantum entanglement}.

Consider a composite quantum system consisting of two or more subsystems.
(For simplicity we consider bi-partite systems in finite dimensional Hilbert 
space, but our results hold for 
any number of particles and in any dimension). Let $\{\Psi \}$ be a set of 
vectors in ${\cal H} =
{\cal H}_1 \otimes {\cal H}_2$. If these vectors are not normalized we can 
consider a set of vectors $\{ \Psi/|| \Psi|| \}$ of norm one in 
${\cal L}$. The set of 
rays of ${\cal H}$ is called the projective Hilbert space  
${\cal P}({\cal H}_1 \otimes {\cal H}_2)$.
If dim${\cal H}_1= N_1$ and dim${\cal H}_2= N_2$, then ${\cal H} \simeq 
{\bf C}^{N_1 N_2}$. The projective Hilbert space is ${\cal P} = 
{\bf C}^{N_1 N_2} - \{0\}/U(1)$ which is a complex manifold of dimension
$(N_1 N_2 -1)$. This can also be considered as a real manifold 
of dimension $2(N_1 N_2 -1)$. Any quantum state 
at a given instant of time can be represented as a point in ${\cal P}$ via
the projection map $\Pi: |\Psi\ra \rightarrow |\Psi\ra \la \Psi|$. 
The evolution of the state vector can be represented by a {\em curve}
 $\Gamma: t \rightarrow |\Psi(t)\ra$ in ${\cal H}$ whose projection 
$\Pi(\Gamma) = {\hat \Gamma}$ lies in ${\cal P}$. Here, smooth mappings 
$\Gamma: [0, t] \rightarrow {\cal L}$ of an interval into a differentiable 
manifold are called smooth curves in the manifold \cite{akp}.

Let ${\cal L}$ be a differentiable manifold and 
$|\Psi\ra \in {\cal L}$. A vector $|v\ra$
is called a {\em tangent vector} to ${\cal L}$ at $|\Psi\ra$  
if there is a smooth
curve passing through $|\Psi\ra$ such that $|v\ra = \frac{|d\Psi\ra}{dt}$.
The {\em tangent space} $T_{|\Psi\ra } {\cal L}$ of ${\cal L}$ at 
$|\Psi\ra$ is the set of all tangent vectors to ${\cal L}$ at
$|\Psi\ra$. The tangent space to a differentiable manifold at the point 
$|\Psi\ra \in {\cal L}$ is a linear space having same dimension as that of
${\cal L}$. What we will prove is that given any multi-particle pure state 
$|\Psi\ra \in {\cal L}$, the tangent vector in infinitesimal time step 
$|d \Psi\ra \in T_{|\Psi\ra } {\cal L}$ is entangled for any generic changes
in the subsystems.

{\it {\bf Infinitesimal change creates entanglement:}}
Here, first I argue that quantum entanglement is necessary for any 
continuous dynamical evolution of multi-particle system.
We know that any continuous, finite time evolution can be thought of as a 
limit of infinite number of sequence of infinitesimal changes. 
Consider a a multi-particle state (say $n$-particle) which is not 
entangled initially. 
Under unitary evolution the initial state evolves as 
$|\Psi(0)\ra= \otimes_{i=1}^n |\psi_i \ra \rightarrow 
|\Psi(T)\ra = U(T)|\Psi(0)\ra$, where
$U(T)= \exp(-iH T)$ and $H$ is the total 
Hamiltonian of the system. The same $U(T)$ can be
obtained from infinitesimal changes via $U(T) = 
{\lim}_{N \rightarrow \infty} (I - iH~T/N)^N$. Now, if $U(t), 
0 < t \le T$ is capable 
of producing entanglement, then the state
can be written as $|\Psi(t)\ra = \sum_{i_1 i_2\cdots i_n} 
C_{i_1 i_2\cdots i_n}(t) |i_1 i_2 \cdots i_n \ra$. There is clearly 
entanglement present at any stage of quantum evolution as well as during 
infinitesimal time steps. The tangent vector $|d\Psi\ra$ is also an 
entangled one.

The surprising thing is that even if $U(t)$ does not produce any entanglement, 
to be able to have a continuous evolution we need quantum entanglement. 
 Suppose we have a composite system that consists of two
subsystems. The state of the combined system is then $|\Psi \ra = 
|\psi_1 \ra \otimes |\psi_2 \ra \in {\cal H}_1 \otimes {\cal H}_2$ , 
where $|\psi_1 \ra \in {\cal H}_1$ and 
$|\psi_2 \ra \in {\cal H}_2$. Now consider the infinitesimal change in the 
state vector $|\Psi\ra$ (i.e., the tangent vector at $\Psi$). 
This is a linear mapping $d: |\Psi \ra \rightarrow |d\Psi \ra$ and can be 
thought of as a derivation at $|\Psi\ra$ on a differentiable manifold 
${\cal L}$. The infinitesimal change in $|\Psi\ra$ is given by
\begin{eqnarray}
|d\Psi\ra  = |\psi_1 \ra \otimes |d\psi_2 \ra +  |d\psi_1 \ra \otimes 
|\psi_2 \ra \in T_{|\Psi\ra } {\cal L}.
\end{eqnarray}
The above state is clearly entangled for generic changes in the 
subsystems $1$ and $2$ as we cannot write $|d\Psi\ra$ as 
tensor product of two infinitesimal changes in the respective Hilbert spaces.
Once we choose coordinates for $|\psi_1 \ra$ and $|\psi_2 \ra$
in ${\cal P}$, then there are no coordinates which can express (1) as 
product states unless $|d \psi_i \ra \propto |\psi_i \ra$, $(i =1,2)$.
But the later corresponds to stationary states, whereby the subsystems do 
not undergo generic change. Now, 
if $(\lambda_0^{(1)}, \lambda_1^{(1)}, ...\lambda_{N_1-1}^{(1)} )$ are 
homogeneous coordinates for $|\psi_1\ra$ and 
$(\lambda_0^{(2)}, \lambda_1^{(2)}, ...\lambda_{N_2-1}^{(2)} )$
are homogeneous coordinates for $|\psi_2\ra$, then the tangent vector can be
written as
\begin{eqnarray}
|d\Psi\ra  = |\psi_1 \ra \otimes \frac{\partial |\psi_2 \ra}{\partial 
\lambda_{i_2}^{(2)} } d\lambda_{i_2}^{(2)} +  
\frac{\partial |\psi_1 \ra}{\partial 
\lambda_{i_1}^{(1)} } d\lambda_{i_1}^{(1)}\otimes |\psi_2 \ra,
\end{eqnarray}
where $(i_1 = 1, 2, \ldots , N_1)$, $(i_2 = 1, 2, \ldots , N_2)$, and 
summation over repeated indices is understood.  
This is also true in any dimension and in multi-particle context. Suppose we 
have a $n$-particle pure product state 
$|\Psi \ra= \otimes_{i=1}^n |\psi_i \ra \in \otimes_{i=1}^n {\cal H}_i $. 
Then the infinitesimal change in the state $|\Psi\ra$ is given by
\begin{eqnarray}
|d\Psi\ra & = &|\psi_1 \ra \otimes |\psi_2 \ra \otimes \cdots 
\otimes |d\psi_n \ra  \nonumber\\
& + & \cdots +  |d\psi_1 \ra \otimes |\psi_2 \ra \otimes \cdots 
\otimes |\psi_n \ra 
\end{eqnarray}
which is again an entangled state for generic changes in the respective 
subsystems. In other words the tangent vector to any 
pure product states is an entangled state. This shows that the infinitesimal 
change is {\em not a local-operation}. It has the ability to create entangled 
states. This is a simple but an important observation that may have many 
ramifications. Here, it is not necessary that all $n$-particles undergo 
infinitesimal change locally. For example, if $(n-1)$-particles out of $n$ 
undergo infinitesimal change locally, then the infinitesimal change in the 
combined state is still entangled. But now the entanglement is 
present between $(n-1)$ and the last one is left out. If we have three 
particles, and the last one does not change, then the infinitesimal change
in the combined state is given by
\begin{eqnarray}
|d\Psi\ra  = (|\psi_1 \ra \otimes |d\psi_2 \ra +  |d\psi_1 \ra \otimes 
|\psi_2 \ra) \otimes |\psi_3\ra.
\end{eqnarray}


Thus, we can say that when the 
state passes through infinitesimal changes entanglement is necessary. This is
because the tangent vector which tells us how the state vector changes is 
typically entangled as given in (3).
This is potentially 
one `hidden power' of quantum entanglement. In any quantum universe, even 
if there is no direct or indirect interaction between constituents, 
mere infinitesimal changes in two or more implies that the infinitesimal 
change in the combined state is entangled! For product states one would think 
that the global change can always be described as local changes. However, our
observation shows, somewhat surprisingly, that whether the composite system is
entangled or non-entangled, {\em  global change cannot always be described as 
local changes}.

We can also quantify how much entanglement is required for a given change 
in the state vector. Geometrically, the change in the state is represented by
a curve whose length is measured in terms of Fubini-Study metric on the 
projective Hilbert space ${\cal P}$ of the quantum system \cite{akp}. 
If we have two 
quantum states that differ infinitesimally, i.e., $|\Psi(t)\ra$ and 
$|\Psi(t+dt)\ra$, then the 
distance between them is
given by $dS^2 = 4 (1 - |\la \Psi(t)\ra |\Psi(t+dt)\ra |^2 ) = 
[\la d\Psi |d\Psi \ra - i(\la \Psi| d\Psi \ra)^2]$. This shows that when the
state changes by $|d\Psi \ra$ the system travels a distance $dS$. Since 
entanglement is necessary for this change, we can say that for a 
distance $dS$ to be traveled by a composite system, 
we need $E(d\Psi/||d\Psi||)$ amount of 
entanglement, where $E(.)$ is will be a measure of entanglement for the 
composite system. For bi-partite system it would be the entropy of the 
any one of the reduced subsystem.

To see the entangling power of infinitesimal change, let us consider two 
identical copies of a (say real) qubit $|\psi \ra = \cos \frac{\theta}{2}
|0\ra + \sin \frac{\theta}{2} |1 \ra$. One can see that the 
infinitesimal change in the two-qubit product state $|\Psi\ra = 
|\psi\ra \otimes |\psi\ra$ is a maximally entangled 
state (unnormalized), i.e., 
 \begin{eqnarray}
|d\Psi\ra =
\sqrt{2} d\theta [\cos \theta |\Psi^+ \ra - \sin \theta |\Phi^-\ra],
\end{eqnarray}
where $|\Psi^+ \ra = \frac{1}{\sqrt 2}(|01\ra + |10\ra)$ and 
$|\Phi^- \ra = \frac{1}{\sqrt 2}(|00\ra - |11\ra)$
are two orthogonal Bell-states. To be precise, the normalized form 
of infinitesimal changed state is a maximally entangled state given by 
$\frac{|d\Psi\ra}{|| d\Psi ||} = 
[\cos \theta |\Psi^+ \ra - \sin \theta |\Phi^-\ra] \in T_{|\Psi\ra } 
{\cal L}$. Thus, to travel a distance $dS$ two real qubits need one ebit of 
entanglement. This implies that 
two identical, non-entangled qubits however far separated, when we look
at the change in the combined state through infinitesimal time steps, then 
the infinitesimal change in the state is a highly non-local state. For example, the quantum 
mechanical correlation in the state $|\Psi(\theta + d\theta) \ra$ (up to
local unitaries) is given by
 \begin{eqnarray}
\la \Psi(\theta + d\theta) |(\sigma. {\bf a} \otimes \sigma.{\bf b}) 
|\Psi(\theta + d\theta) \ra & = & \la \sigma. {\bf a} \ra 
\la \sigma. {\bf a} \ra  \nonumber\\
& + & E({\bf a}, {\bf b}) d\theta^2,
\end{eqnarray}
where $E({\bf a}, {\bf b})= - {\bf a}.{\bf b}$ is standard quantum mechanical
correlation is a maximally entangled state. This means 
if one looks at change in infinitesimal steps, one may observe violation 
of Bell's inequality even for product states.

Note that the infinitesimal change of the global state cannot be a bi-local 
infinitesimal operation, i.e., an operation taking $|\Psi\ra \rightarrow 
|\Psi'\ra = 
(d \otimes d)|\Psi\ra$ is an impossible one. This violates the norm 
preservation. For bi-partite systems this would 
mean $|\psi_1 \ra \otimes |\psi_2 \ra \rightarrow 
|d \psi_1 \ra \otimes |d\psi_2 \ra$ which cannot happen. We can prove this by 
contradiction. Suppose we have the 
mapping $f: |\Psi\ra \rightarrow |\Psi' \ra = (d \otimes d)|\Psi\ra$. Then we
have $\la \Psi|\Psi'\ra = \la \psi_1|d\psi_1\ra \la \psi_2|d\psi_2\ra$.
For any normalized state $|\psi\ra$ we must have $\la \psi |d\psi \ra$ as
a purely imaginary number. This implies that on lhs we have 
$\la \Psi|\Psi' \ra$ which is a purely 
imaginary number and on rhs we have product of two purely imaginary numbers 
which is a real number.
Since this cannot hold, there is no bi-local infinitesimal changes. The
proof can be generalized for more than two subsystems.
However, if one of of the subsystem does not undergo infinitesimal change 
then it is possible to satisfy the norm preservation or isometric evolution.  
This means we can have $|\Psi\ra \rightarrow  
(d \otimes I)|\Psi\ra$ and $|\Psi\ra \rightarrow
(I \otimes d)|\Psi\ra$ but not $|\Psi\ra \rightarrow (d \otimes d)|\Psi\ra$.

{\it {\bf Infinitesimal change and reduced dynamics:}}
In quantum information theory entanglement plays a dual role: sometimes 
it acts as a perfect quantum channel and sometimes also acts as a noisy 
channel. When we describe a quantum operation ${\cal E}$ acting on a system, 
we can always imagine ${\cal E}$ as a unitary evolution on a combined system 
(system + ancilla) and then tracing over the ancilla. If $\rho$ is the state 
(pure or mixed) of the system and  
$\rho \rightarrow {\cal E}(\rho)$, then ${\cal E}(\rho)=  
{\rm tr}_2 (U \rho \otimes \sigma U^{\dagger} )$.  This unitary version of 
quantum operation always produces
an entangled state which in effect amounts to passing the system through a 
noisy channel \cite{nc}. Now one may ask since the infinitesimal change in the
combined product state creates an entangled state what kind of noise does 
that introduce for reduced dynamics. First, we note that the global 
infinitesimal change is not a unitary evolution \cite{note}. 
 The infinitesimal change of the combined state 
is precisely the operator represented as 
$(I_1 \otimes d_2 + d_1 \otimes I_2)$, which by itself is not unitary.
But still we can ask what is the reduced dynamics of any one of the subsystem.
It can be verified that when 
$|\Psi \ra \rightarrow |d\Psi\ra  = |\psi_1 \ra \otimes |d\psi_2 \ra +  
|d\psi_1 \ra \otimes |\psi_2 \ra$, then the first subsystem undergoes the
evolution as given by 
 \begin{eqnarray}
|\psi_1\ra \la \psi_1| &\rightarrow & {\cal D}(|\psi_1\ra \la \psi_1|) = 
{\rm tr}_2 (|d \Psi\ra \la d \Psi| ) = |d\psi_1\ra \la d\psi_1| \nonumber\\
 & + & (|\psi_1\ra \la d\psi_1| - 
|d\psi_1\ra \la \psi_1| ) \la \psi_2|d\psi_2 \ra \nonumber\\
& + & |\psi_1\ra \la \psi_1| 
 \la d\psi_2|d\psi_2 \ra.
\end{eqnarray}
Here, ${\cal D}$ may be thought of as the quantum channel arising from 
global infinitesimal changes. 
This clearly shows that when $|\Psi \ra \rightarrow |d\Psi\ra$, 
then $|\psi_1 \ra$ does not simply go to $|d\psi_1 \ra$, rather there are
additional contributions coming due to entangled nature of infinitesimal change
of the combined state. Similarly, we can see that the second subsystem 
undergoes the evolution given by
 \begin{eqnarray}
|\psi_2\ra \la \psi_2| &\rightarrow & {\cal D}(|\psi_2\ra \la \psi_2|) = 
{\rm tr}_1 (|d \Psi\ra \la d\Psi| ) = |d\psi_2\ra \la d\psi_2| \nonumber\\
 & + & (|\psi_2\ra \la d\psi_2| - 
|d\psi_2\ra \la \psi_2| ) \la \psi_1|d\psi_1 \ra \nonumber\\
& + & |\psi_2\ra \la \psi_2| 
 \la d\psi_1|d\psi_1 \ra.
\end{eqnarray}
This shows that when $|\Psi \ra \rightarrow |d\Psi\ra$, 
then $|\psi_2 \ra$ transforms to $|d\psi_2 \ra$ along with noise terms. 
Eqs(7) and (8) clearly show the entangling nature of infinitesimal change. 
If it has no ability to create entanglement, then 
$|\psi_1 \ra$ would have gone to $|d\psi_1 \ra$
and $|\psi_2 \ra$ would have gone to $|d\psi_2 \ra$ under global infinitesimal
changes.

{\it {\bf Implication for quantum computing:}}
Usual quantum computation paradigm involves preparation of initial logical 
states and application of sequence of unitary evolution operators 
(prescribed by a particular quantum mechanical algorithm) and then finally 
reading out the desired answer \cite{nc}. 
In this context an important question has been
whether linear superposition alone is sufficient to have the required speed-up
or we need quantum entanglement--the weirdest feature of quantum world. 
Though the existing quantum algorithms such as
Deustch-Jozsa \cite{collin}, Grover \cite{bp} and Shor \cite{lp} 
require quantum entanglement it is not clear
whether in general entanglement is the key for quantum speed-up \cite{jl} .  
In particular, there has been debates in NMR implementation of quantum 
algorithms as to what gives the power to quantum computers if there is no
entanglement generated during computation \cite{sam}.

I show that even though the initial state of 
$n$-qubit register is a product state, even though all the $n$-qubits during
computation are product states, entanglement is necessary for quantum 
computation. To see this clearly, consider the initial state of $n$-qubit 
state prepared in equal superpositions (which is a product state) 
$|\Psi_0 \ra = \frac{1}{\sqrt 2^n} \sum_{i=0}^{2^n -1} |x_i \ra$.
At any stage of the computation (say $k$th step) we can write the $n$-qubit
state generically as 
\begin{eqnarray}
|\Psi_k \ra = \bigg[ U_k U_{k-1}  \cdots U_1 \bigg] |\Psi_0\ra
\end{eqnarray}
Suppose that the unitary operators $U_k$'s are such that at any stage of the 
computation there is no entanglement generated. Can we say that entanglement 
has no role to play? The answer is no.
From our earlier observation, we know that even if a multi-particle
state is a product state, infinitesimal change in the multi-particle product
state is an entangled state. Thus, if we look at infinitesimal changes during
quantum computation we will see that the infinitesimal change in the $n$-qubit 
registrar is indeed entangled. 
For example, the state at $k$th step can be written as 
\begin{eqnarray}
|\Psi_k \ra = \bigg[ u_1(k) \otimes u_2(k) \otimes  \cdots u_n(k) \bigg]
|\Psi_0 \ra
\end{eqnarray} 
where each of these $u_i$'s are some local unitaries acting on single qubit 
Hilbert space ${\cal H}^2$.
But the infinitesimal change in the above state 
$|d \Psi_k \ra \in T_{|\Psi_k \ra } {\cal L}$ is given by 
\begin{eqnarray}
|d\Psi_k \ra & = & \bigg[du_1(k) \otimes u_2(k) \otimes  
\cdots \otimes u_n(k) 
+ \cdots \nonumber\\ 
& + &  u_1(k) \otimes u_2(k) \otimes  \cdots \otimes du_n(k) \bigg]
|\Psi_0 \ra
\end{eqnarray} 
which is an entangled state.
This shows that the weirdest feature of 
quantum world plays its role in every computation in a very subtle way.
This is truly a `hidden power' of quantum entanglement in quantum 
computation. It is hidden because, we do not look at infinitesimal steps; 
we always consider finite time steps. Quantum entanglement is 
necessary for any evolution 
(entangling or not) and hence for quantum computation.
We can say that {\em during quantum computation possible directions in 
which one can 
pass through a $n$-qubit register state is guided by entanglement.}

One may ask does entanglement also plays any role when mixed state are
involved during quantum computation? This is exactly the case when one deals 
with NMR implementations. There one typically encounters pseudo-pure states
which comes as a convex combination of a random mixture and a pure state.
For $n$-qubits this is given by
$\rho = (1 -\epsilon) \frac{I}{2^n} + \epsilon |\Psi_0 \ra \la \Psi_0|$,
where $\epsilon$ is the purity parameter.
After application of sequence of unitary operators during certain 
computation the state changes as
$\rho \rightarrow U \rho U^{\dagger} = 
(1 -\epsilon) \frac{I}{2^n} + \epsilon |\Psi \ra \la \Psi |$.
It has been claimed that the states produced thus are still
not entangled even though the pure state component $|\Psi \ra =  U|\Psi_0\ra$ 
is entangled. However, now we see how does entanglement play a role
during quantum computation? What we say is 
that even though the the pure state component $|\Psi \ra$ is 
not entangled, the infinitesimal change in $\rho $, i.e., $d\rho$ 
indeed is entangled. This is because $d\rho = \epsilon [|\Psi \ra \la d\Psi | 
+  |d\Psi \ra \la \Psi ]$ is an entangled one.
Thus any continuous evolution of $\rho$ does require quantum entanglement.

Our observation not only applies to multi-particle pure and pseudo-pure states 
but to any mixed states as well. Consider a separable multi-particle 
(again for simplicity say bi-partite) mixed state given by
\begin{eqnarray}
\rho = \sum_i p_i \rho_i^{(1)} \otimes \rho_i^{(2)},
\end{eqnarray} 
where $\rho_i^{(1)} \in {\cal B}({\cal H}_1)$ and $\rho_i^{(2)} 
\in {\cal B}({\cal H}_2)$ are pure state components of subsystems
$1$ and $2$, respectively with $p_i > 0, \sum_i p_i =1$. We can show that 
even this separable state when evolves in time, the infinitesimal change in 
the state is an entangled one. The infinitesimal change in $\rho$ is given by
\begin{eqnarray}
d\rho = \sum_i p_i [\rho_i^{(1)} \otimes d\rho_i^{(2)} + 
d\rho_i^{(1)} \otimes \rho_i^{(2)}].
\end{eqnarray} 
To prove that $d\rho$ is entangled, let us assume that it is separable
and then arrive at a contradiction. If $d\rho$ is separable then there must be
a decomposition such that we can write this as
\begin{eqnarray}
d\rho = \sum_i w_i d \sigma_i^{(1)} \otimes d\sigma_i^{(2)} 
\end{eqnarray} 
for some pure components $\sigma_i^{(1)} \in {\cal B}({\cal H}_1)$ 
and $\sigma_i^{(2)} \in {\cal B}({\cal H}_2)$ and $w_i > 0, \sum_i w_i =1$. 
This is the only form consistent with separability because the state is 
classically correlated and there are infinitesimal changes for both the 
subsystems. This then implies that if $d\rho$ is separable we have
\begin{eqnarray}
(I \otimes d + d \otimes I) \rho = (d \otimes d) \rho
\end{eqnarray} 
which cannot be satisfied for arbitrary $\rho$.
Alternately, if we look at the reduced changes then
from (13) we have ${\rm tr}_2(d\rho) = \sum_i p_i d\rho_i^{(1)}$ and from 
(14) we have ${\rm tr}_2(d\rho) = 0$ which is a clear contradiction. Here, we 
have used the fact that  ${\rm tr}_2(d\rho_i^{(2)}) =0$ and  
${\rm tr}_2(d\sigma_i^{(2)}) =0$ which is true for any pure state 
density operators. Thus, the infinitesimal change in any multi-particle 
separable density operator is an entangled one. 
This shows that any multi-particle {\em quantum evolution} be it pure or 
mixed does require quantum entanglement.
One may ask since any classical computer state
can be written as a separable state would that have the same
behavior? The answer is no, because to have entangled tangent vector we need
derivative behavior and tensor product structure on the linear space.
Also one may ask if we have 
ordinary probabilistic description for a system comprising two subsystems
would we say that non-product states are necessary
for any probabilistic evolution as well? The answer will depend on whether we
have tensor product structure on a linear space as a form of description.

{\bf Conclusion:}
We have shown that entanglement is necessary for 
any quantum evolution. In short any generic change in a quantum universe 
does require entanglement.
Geometrically, given a manifold of multi-particle quantum states if there 
are changes in two or more subsystems then the 
tangent space is an entangled manifold. 
Since the tangent space vectors tell the state-space vectors how 
to change 
so our result tells us something deep about motion or 
change in general.
Also we have shown that infinitesimal change cannot be a bi-local operation. 
We have studied the reduced dynamics of the subsystem under infinitesimal 
operation.
This has immediate implication in quantum 
computation, where one can argue that even though there is no entanglement 
generated during any stage of computation, the evolution of 
multi-qubit state is guided by entanglement.
This result applies to pure state, pseudo-pure state and mixed state 
implementations as well. 
Though the result 
of this paper may appear simple, it is nevertheless non-trivial. 
We hope that this observation will unfold many other results in quantum 
theory and in the fast growing field of
quantum information theory.

{\it Acknowledgment:} I thank S. L. Braunstein for useful comments. Also I 
would like to thank R. Jozsa for his remarks. Comments from A. K. Rajagopal, 
P Gralewicz and R. Garisto are highly appreciated.\\

\small {\sf Email:akpati@iopb.res.in}




\begin{thebibliography}{99}
\bibitem{note} The infinitesimal local unitary evolution is given by
$U_1 \otimes U_2 = I_1 \otimes I_2 - (I_1 \otimes d_2 + d_1 \otimes I_2) + 
d_1 \otimes d_2$,
where $I_1$, $I_2$ are identity matrices and $d_1$ and $d_2$ are infinitesimal 
changes acting on ${\cal H}_1$ and 
 ${\cal H}_2$, respectively. Here, first and third terms are local operators 
but the second term is not.
\end{thebibliography}
\end{document}